# Is the assumption of a special system of reference consistent with Special Relativity? Definitely yes, special approaches show it!

Bernhard Rothenstein[1], Stefan Popescu[2] and George J. Spix[3]

1) Politehnica University of Timisoara, Physics Department, Timisoara, Romania, bernhard_rothenstein@yahoo.com
2) Siemens AG, Erlangen, Germany, stefan.popescu@siemens.com
3) BSEE Illinois Institute of Technology, USA, gjspix@msn.com

**Abstract.** *We compare the results obtained by interpreting some fundamental relativistic experiments from the point of view of two alternative theories: Einstein's special relativity theory and the Lorentz-Poincare theory admitting the existence of a special reference frame relative to which we can measure absolute velocities. The experiments that we consider here are the time dilation, the radar echo, the Doppler Effect and the radar detection. The most important result is that both theories lead to the Lorentz-Einstein transformations for the space-time coordinates of the same event. This conclusion enables us to assert that the anisotropy in the light propagation fades away in the case of these particular experiments.*

## 1. Introduction

There are two competitive approaches to the relativity theory, which differ in essence in the aspect of admitting or denying the possibility to measure the absolute velocity of particles:

1. In the first approach the concepts of "really resting" and "really moving" are meaningless (Einstein).
2. In the second approach there is indeed a state of real rest defined by the "ether" even tough the laws of physics conspire by preventing us to identify it experimentally (Lorentz-Poincare)[1].

This problem has many facets (convention in the synchronization of distant clocks, immeasurability of the one-way speed of light, anisotropy of light propagation) being covered largely in the literature[2].

The first approach starts by stating Einstein's principle of relativity:
- the laws of physics are the same in all inertial reference frames in relative motion,
- the light propagates in all directions with the same speed *c* relative to all inertial reference frames and is not influenced by the state of motion or rest of the light source.

The second approach starts by stating that:

- some medium exists (called "ether") as well as an absolute reference frame attached to it, in which the light propagates in free space in a straight line with the same speed *c* in all directions,
- in any other inertial reference frame the two-way propagation velocity of light is invariant.

Others state:
- all good clocks can be used to measure time, independently of the periodic physical phenomena they are built upon,
- time is measured in the same way in all inertial reference frames, i.e., if a particular clock can be used to measure time in the "rest system", a similar clock can be used to measure time in any other moving inertial frame
- a speed limit exists in the rest "system".[3]

Let K(XOY) be the rest frame relative to which we measure the absolute speed of particles. For Einstein's theory it represents an arbitrarily chosen inertial reference frame. According to Lorentz's theory it is a unique reference frame (discovered by chance i.e. don't ask how) relative to which we are able to measure absolute speeds. The light propagates relative to it, in free space, in all directions with the same speed *c* independently of the speed of the source. Let K'(X'O'Y') be another inertial reference frame. In Einstein's special relativity it moves with constant speed *u* relative to K in the positive direction of the common OX(O'X') axes, the light propagating relative to it under the same conditions as in K. In the Lorentz approach K' moves with absolute velocity *u* relative to K, the light propagating relative to K' with different speeds in different directions. Considering only the common OX(O'X') axes, the light propagates in its positive direction with speed $c_+$ but with speed $c_-$ in its negative direction. The $c_+$ and $c_-$ represent one-way velocities of light. Consider the following experiment performed in K', having the properties imposed by Lorentz. Let $L_0$ be the proper length of a rod located along the O'X'. It is at rest relative to K'. A source of light at rest in K' is located at its left end with a mirror being located at its right end. The source emits a light signal, in the positive direction of the O'X' axis, that returns back, after reflection, to the left end. The trip of the light signals lasts:

$$\frac{L_0}{c_+} + \frac{L_0}{c_-}. \tag{1}$$

All experiments performed in order to measure the speed of light may only determine its two-way speed *c*. The trip of the light signal expressed as a function of the two-way light speed lasts:

$$\frac{2L_0}{c}. \tag{2}$$

Combining (1) and (2) we obtain that the velocities defined above are related by

$$\frac{2}{c} = \frac{1}{c_+} + \frac{1}{c_-}. \tag{3}$$

If we arbitrary chose a value for $c_+$ then from (3) we get the corresponding value for $c_-$. Strange enough, if we chose $c_+ \to \infty$ we obtain $c_- = \frac{c}{2}$.

As usually the corresponding axes of the two frames are parallel to each other, the OX(O'X') axes are common and at the origin of time the origins O and O' are located at the same point in space. In the reference frame K we have the rest observers $R_i(x_i,0)$ and their wrist watches $C_i(x_i,0)$ located at the different points $x_i$ on the OX axis. In both theories we can synchronize their clocks following a synchronization procedure proposed by Einstein. In the reference frame K' we find the observers $R'_i(x'_i,0)$ and their wristwatches $C'_i(x'_i,0)$. In Einstein's approach we can also synchronize them following his procedure. In the Lorentz approach Einstein's synchronization procedure doesn't work and consequently other, less or more feasible synchronization procedures are proposed. In our approach an important part is played by the observers $R_0(0,0)$ and $R'_0(0,0)$ located at the corresponding origins O and O' and by their wristwatches $C_0(0,0)$ and $C'_0(0,0)$. An initialization procedure can ensure that the clocks $C_0(0,0)$ and $C'_0(0,0)$ read a zero time when they are located at the same point in space.

Consider a given physical object located somewhere in space. In order to characterize it the physicists invent physical quantities. Performing experiments the physicists discover physical laws that establish a relationship between different physical quantities. The physical object can be studied by observers from K and by observers from K' as well. The problem of relativity is to find out a relationship between the same physical quantity, introduced to characterize the same physical object as measured by observers from K and respectively K'.

The considerations we've made above suggest that the two theories should lead to the same results if we follow scenarios in which:
- ***Clock synchronization is not performed nor in K neither in K'*** because the anisotropy in the light propagation has no occasion to manifest itself,

- *Clock synchronization is performed in K but not in K'* because both theories agree that clock synchronization could be performed in K (following Einstein's clock synchronization procedure) but not in K',
- *Two-way experiments are performed in both reference frames* and the observers from the two frames operate exclusively with the two-way speed of light considering it a true relativistic invariant.

We will revise the following particular experiments:
- *The time dilation*[4]**:** An experiment that requires clock synchronization in K but not in K',
- *The radar echo*[4]**:** A two-way experiment that doesn't require clock synchronization in the reference frames involved,
- *The relativistic Doppler Effect*[4]**:** An experiment that doesn't require clock synchronization in any of the reference frames involved,
- *The radar detection*[5]: An experiment that has a two-way character in both frames.

**2. Time dilation**
Many textbooks[6] use the light clock shown in figure 1 for deriving the time dilation formula. We use it for showing that it relates a proper time interval to a non-proper time interval:
- the first is measured in K' as a difference between the readings of the same clock $C'_0$
- the second is measured as a difference between the readings of two distant clocks of the K frame, synchronized following a clock synchronization procedure proposed by Einstein.

Figure 1.a shows how the clock $C'_0$ located on the mirror $M'_1$ measures the period $T'$ of the light clock at rest in K'. The period $T'$ of the light clock equates the reading $t'$ of $C'_0$ and as a result of a two-way experiment we have obviously:
$$t' = T' = \frac{2d}{c} \qquad (4)$$

$c$ representing the two-way speed of light and $d$ the invariant distance between the mirrors of the light clock ($M'_1$ and $M'_2$).

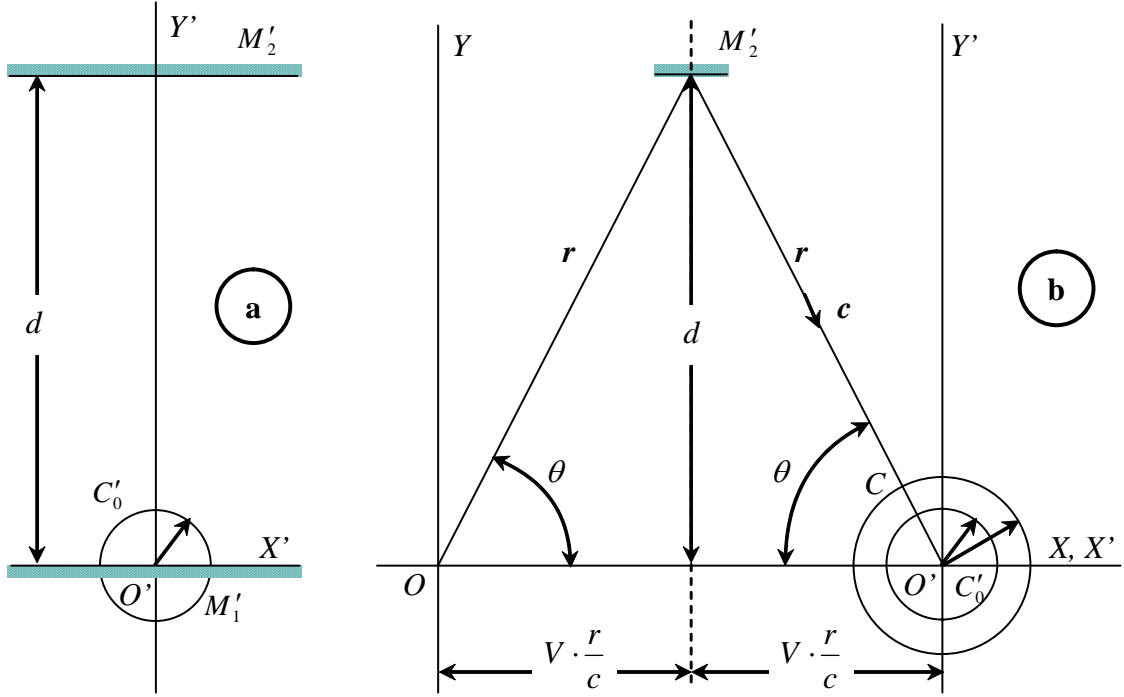

*Figure 1. The two clocks of the reference frames K'(X'O'Y') and K(XOY)*

Figure 1b shows the same experiment as viewed from the K frame. Clock $C'_0$ is reading $t' = \dfrac{2d}{c}$ when located in front of a clock $C(x = 2V\dfrac{r}{c}, y = 0)$ of the K frame. The clock $C$ is synchronized as proposed by Einstein with the clock $C_0$ that reads $t = T = \dfrac{2r}{c}$. Here $T$ represents the period of the light clock but also the reading of the clock $C$. When the reflection of light takes place on the mirror $M'_2$ its polar coordinates are $(r, \theta)$. Pythagoras' theorem leads to

$$t = \frac{t'}{\sqrt{1 - \dfrac{V^2}{c^2}}} \tag{5}$$

and to

$$T = \frac{T'}{\sqrt{1 - \dfrac{V^2}{c^2}}} \tag{6}$$

By definition $T' = t' - 0$ represents a **proper time interval** measured as a difference between the readings of the same clock $C'_0$. Its measurement does not involve clock synchronization being the result of a two-way experiment. $T = t - 0$ represents a **non-proper time interval**, measured as a difference between the readings of two distant clocks $C$ and $C_0$. Both approaches to special relativity consider that (5) and (6) are correct because they don't violate the postulates on which they are based.

### 2. The radar echo

The radar echo experiment performed in the K frame involves a source of light $S_0$ located at O, the observer $R_0$ and his wristwatch $C_0$. A mirror M is moving with speed $u$ in the positive direction of the OX axis being located in front of $S_0$ when $C_0$ reads $t=0$. When $C_0$ reads $t=0$ the source $S_0$ emits a first light signal that is instantly received and reflected back by the mirror. A second light signal is emitted when $C_0$ reads $t_e$ arriving at the mirror when a clock $C$ of the K frame located just in front of the mirror reads $t$. The reflected light signal arrives at $R_0$ when his wristwatch reads $t_r$. We write the obvious equations that expresses in two different ways the distances travelled by the mirror and by the direct and the reflected rays respectively as

$$u_+ t = c(t - t_e) \tag{7}$$

and

$$u_+ t = c(t_r - t). \tag{8}$$

Eliminating $t$ between (7) and (8) we obtain

$$\frac{t_r}{t_e} = \frac{1 + \dfrac{u_+}{c}}{1 - \dfrac{u_+}{c}}. \tag{9}$$

Considering that $T_e = t_e - 0$ represents the period at which the source emits successive light signals and that $T_r = t_r - 0$ represents the period at which the reflected signals are received back, we can present (9) as

$$\frac{T_r}{T_e} = \frac{1 + \dfrac{u_+}{c}}{1 - \dfrac{u_+}{c}} \tag{10}$$

We underline that the time intervals $T_e$ and $T_r$ are time intervals measured as differences between the readings of the same clock $C_0$ and therefore no clock synchronization is involved.

We have followed so far the first step of a hint that states: "*When you have a situation which you don't fully understand find a new reference frame in which you do understand it. Examine it in this new reference frame. Then translate your understanding that you gained in the new frame back into the language of the old frame.*"[7] Following the second step switch over to the K' frame in the Lorentz approach. The observer $R'_0$ knowing the properties of his rest frame and performing the radar echo experiment obtains

$$\frac{T'_r}{T'_e} = \frac{1+\dfrac{u'_+}{c_+}}{1-\dfrac{u'_+}{c_-}} \qquad (11)$$

where $T'_r$ and $T'_e$ represent time intervals measured as differences between the readings of the same clock $C'_0$ initialized with clock $C_0$ to read both a zero time when they are located at the same point in space. Because $T_e, T'_e$ and $T_r, T'_r$ are related to each other in the same way i.e. $\dfrac{T_r}{T_e} = \dfrac{T'_r}{T'_e}$ we conclude that

$$\frac{1+\dfrac{u_+}{c}}{1-\dfrac{u_+}{c}} = \frac{1+\dfrac{u'_+}{c_+}}{1-\dfrac{u'_+}{c_-}}. \qquad (12)$$

If we choose

$$c_+ = \frac{c}{1+|s|} \qquad (13)$$

then (3) imposes for $c_-$ a value

$$c_- = \frac{c}{1-|s|} \qquad (14)$$

with the synchronization factor $|s| \leq 1$.

Combining the equations derived above we obtain the following relationships between the isotropic and the anisotropic speeds defined above

$$u_+ = \frac{u'_+}{1-|s|} \qquad (15)$$

$$u'_+ = \frac{u_+}{1+|s|}. \qquad (16)$$

Considering the same experiment this time with the mirror approaching the source with speed $u_-$ ($u_+ = u_-$) we obtain in a similar way

$$u_- = \frac{u'_-}{1-|s|u'_+ c^{-1}} \qquad (17)$$

$$u'_- = \frac{u_-}{1+|s|u_- c^{-1}} \qquad (18)$$

The important conclusion here is that the radar echo experiment leads to the invariant combinations of $c$ and $u_+$ and respectively of $\frac{u'_+}{c_+}$ and $\frac{u'_+}{c_-}$. This is why we can say that in the case of the radar echo experiment the anisotropy in the propagation of light fades away.

### 3. The Doppler Effect

A Doppler Effect experiment involves in one of its simplest variants a stationary source of light $S_0$ that emits successive light signals at constant time intervals $T_e$ measured as a difference between the readings of the same clock $C_0$ being by definition a proper time interval. One observer $R'_0$ moving with constant speed $u_+$ relative to K receives these light signals at constant time intervals $T'_r$ measured as a difference between the readings of his wrist watch $C'_0$ being by definition a proper time interval as well. The two periods are related in Einstein's special relativity by

$$\frac{T'_r}{T_e} = \sqrt{\frac{1+\frac{u_+}{c}}{1-\frac{u_+}{c}}} \qquad (19)$$

but in the Lorentz approach by

$$\frac{T'_r}{T_e} = \sqrt{\frac{1+\frac{u'_+}{c_+}}{1-\frac{u'_+}{c_+}}} \qquad (20)$$

If $R'_0$ changes the direction in which he moves, then in Einstein's approach we have

$$\frac{T'_r}{T_e} = \sqrt{\frac{1-\dfrac{u_-}{c}}{1+\dfrac{u_-}{c}}} \qquad (21)$$

but in the Lorentz approach we have

$$\frac{T'_r}{T_e} = \sqrt{\frac{1-\dfrac{u'_-}{c_+}}{1+\dfrac{u'_-}{c_+}}} \; . \qquad (22)$$

Because the postulates on which these two approaches are based don't conflict in the case of the relativistic Doppler Effect we conclude that all the equations derived above shall hold exactly and shall lead to the same results. Convincingly we can prove it by expressing the right side of (21) as a function of physical quantities measured in K' or expressing the right hand side of (22) as a function of physical quantities measured in K via the corresponding transformation equations derived above. This proves that the combination of physical quantities under the square root in (21) and (22) is a true invariant.

**4. Radar detection of the space-time coordinates of the same event**

The radar detection procedure of the space time coordinates of a distant event performed in the K frame, involves the observer $R_0$ and his wrist watch $C_0$. In order to detect the space-time coordinates of an event he emits a light signal towards the point where the event $E(x,0,t)$ takes place when $C_0$ reads $t_e$. The light signal is reflected by a mirror $M(x,o)$ located where the event takes place when a further clock $C(x,o)$ of the K frame synchronized with clock $C_0$ reads $t$. The reflected light signal returns to O when clock $C_0$ reads $t_r$. We write the obvious equations

$$x = c(t - t_e) \qquad (23)$$

and

$$x = c(t_r - t) \qquad (24)$$

which express in two different ways the distances travelled by the emitted and respectively the reflected light signals. Eliminating $t$ between (23) and (24) we obtain

$$x = \frac{c}{2}(t_r - t_e) \qquad (25)$$

and

$$t = \frac{1}{2}(t_r + t_e) \qquad (26)$$

The observer $R'_0$ devising the same experiment considers that both theories should lead to the same result because it has a two-way character and doesn't involve clock synchronization. So, he declares that the first postulate works in both approaches, therefore (24) and (25) are reading from his point of view

$$x' = \frac{c}{2}(t'_r - t'_e) \tag{27}$$

$$t' = \frac{1}{2}(t'_r + t'_e) . \tag{28}$$

Figure 2 illustrates an attempt by $R_0$ and $R'_0$ to detect the space-time coordinates of the same event $E(x,0,t)$ and $E(x',0,t')$ using the radar procedure as displayed by a classical space-time diagram. The events involved are

- **1(0,0,$t_e$)** "Source S(0,0) emits a radar signal when $C_0$ reads $t_e$."
- **E(x,0,t)** "The radar signal arrives at the point where the detected event takes place being instantly reflected back."
- **2(0,0,$t_r$)** "The reflected light signal returns to $R_0$."

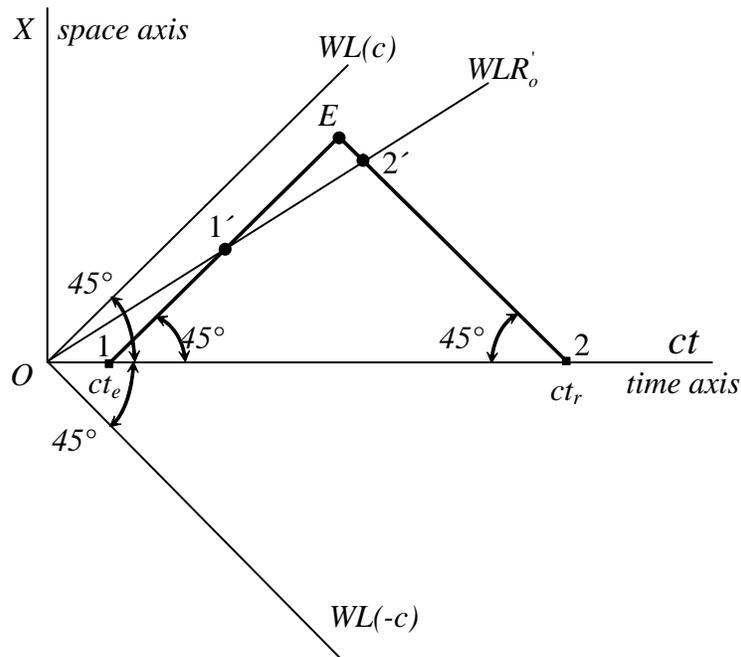

**Figure 2.** *Radar detection on the space-time diagram*

The space-time diagram displays the world line of observer $R'_0$ ($WLR'_0$) making an angle $\alpha$ with the time axis of the diagram given by

$$\tan\alpha = \frac{u}{c}. \tag{29}$$

The intersection of ($WLR'_0$) with the emitted and the reflected light signals generates the events **1'** and respectively **2'**, characterized in K' by the time coordinates $t'_e$ and respectively $t'_r$. We consider that the radar detection procedure is associated with two successive Doppler Effect experiments. In the first one we have

$$t'_e = t_e \sqrt{\frac{1+\dfrac{u_+}{c}}{1-\dfrac{u_+}{c}}} \tag{30}$$

whereas in the second one we have

$$t_r = t'_e \sqrt{\frac{1+\dfrac{u_+}{c}}{1-\dfrac{u_+}{c}}}. \tag{31}$$

Expressing the right hand sides of (27) and (28) as a function of physical quantities measured in the K frame via (29) and (30) we obtain

$$x' = \frac{x-Vt}{\sqrt{1-\dfrac{V^2}{c^2}}} \tag{32}$$

and

$$t' = \frac{t-\dfrac{V}{c^2}x}{\sqrt{1-\dfrac{V^2}{c^2}}} \tag{33}$$

recovering the same result obtained by Abreu[8] when starting from different premises.

In a consequent Lorentz approach[9] the space coordinates of the same event transform in accordance with (32). An important conclusion is that in the case of the radar detection (a two-way experiment) both theories lead to the Lorentz-Einstein transformation, an unexpected result showing that it is not correct to argue like "Poincare versus Einstein" but it is better to state "Poincare hand in hand with Einstein".

**Conclusions**

Ungar shows[2] in a rigorous but not very simple way that in the case of the relativistic Doppler Effect anisotropy fades away.

Furthermore we have shown that the anisotropy fades away also in the case of the relativistic effects in which synchronized clocks are involved only in the K frame (time dilation) or no synchronization is involved in any frame (radar echo and radar detection). Both relativistic approaches lead to the same results concerning the transformation equations of the space-time coordinates. A difference still persists concerning the symmetry.

The surprising result that both theories lead to the Lorentz-Einstein transformation for the space-time coordinates of the same event suggests that they should lead to the same results in all other cases.